\makeatletter
\declare@file@substitution{revtex4-1.cls}{revtex4-2.cls}
\makeatother

\documentclass[twocolumn]{aastex631}

\usepackage{graphicx}
\usepackage{dcolumn}
\usepackage{bm}
\usepackage{epsf}
\usepackage{epsfig}
\usepackage{verbatim}
\usepackage{amsmath}
\usepackage{amsfonts}
\usepackage{ifthen}
\usepackage[normalem]{ulem}
\usepackage[flushleft]{threeparttable}

\newcommand{\gama}{{$\gamma$}}
\newcommand{\eV}{{\mbox{ eV}}}
\newcommand{\keV}{{\mbox{ keV}}}
\newcommand{\GeV}{{\mbox{ GeV}}}
\newcommand{\kpc}{{\mbox{ kpc}}}
\newcommand{\eg}{{\emph{e.g.},}}
\newcommand{\cm}{{\mbox{ cm}}}
\newcommand{\km}{{\mbox{ km}}}
\newcommand{\se}{{\mbox{ s}}}
\newcommand{\Myr}{{\mbox{ Myr}}}

\newcommand{\GHz}{{\mbox{ GHz}}}
\newcommand{\MHz}{{\mbox{ MHz}}}
\newcommand{\muG}{{\mu\mbox{G}}}
\newcommand{\coma}{{\, ,}}
\newcommand{\fin}{{\, .}}
\newcommand{\myv}{{\upsilon}}

\newcommand{\myEps}{{\varepsilon}}
\newcommand{\myfp}{{q}}

\begin{document}

\title{Fermi-bubble bulk and edge analysis  reveals dust, cooling breaks \\ and cosmic-ray diffusion, facilitating a self-consistent model}

\author{Uri Keshet}
\affiliation{Physics Department, Ben-Gurion University of the Negev, POB 653, Be'er-Sheva 84105, Israel}
\email{keshet.uri@gmail.com}
\author{Ilya Gurwich}
\affiliation{Qedma Quantum Computing, Tel-Aviv, Israel}
\affiliation{Physics Department, NRCN, POB 9001, Be'er-Sheva 84190, Israel}
\author{Assaf Lavi}
\affiliation{Physics Department, Ben-Gurion University of the Negev, POB 653, Be'er-Sheva 84105, Israel}
\author{Dina Avitan}
\affiliation{Computer Science Department, Ben-Gurion University of the Negev, POB 653, Be'er-Sheva 84105, Israel}
\author{Teodor Linnik}
\affiliation{Computer Science Department, Ben-Gurion University of the Negev, POB 653, Be'er-Sheva 84105, Israel}

\begin{abstract}
The full, radio to $\gamma$-ray spectrum of the Fermi bubbles is shown to be consistent with standard strong-shock electron acceleration at the bubble edge, without the unnatural energy cutoffs and unrealistic electron cooling of previous studies, if the ambient interstellar radiation is strong; the $\gamma$-ray cooling break should then have a microwave counterpart, undetected until now.
Indeed, a broadband bubble-edge analysis uncovers a pronounced downstream dust component, which masked the anticipated $\sim35$ GHz spectral break,
and the same overall radio softening consistent with Kraichnan diffusion previously reported in $\gamma$-rays.
A self-consistent bulk and edge model implies a few Myr-old bubbles, with fairly uniform radiation fields and enhanced magnetization near the edge.
\end{abstract}

\section{\bf I\lowercase{ntroduction}}
The bipolar Fermi bubbles (FBs), emanating from the center of the Milky Way \citep{Baganoffetal03, Blandhawthorncohen03} and extending out to $|b|\gtrsim 50^\circ$ latitudes, each presents as a fairly uniform \gama-ray lobe \citep{DoblerEtAl10, SuEtAl10}, demarked by an X-ray shell \citep{Keshetgurwich18}, and coincident with microwave emission \citep{Dobler12, PlanckHaze13} which brightens to a haze \citep{Finkbeiner04} near the Galactic center.
Their morphology, X-ray shells \citep{Keshetgurwich18}, and bulk \citep[volume-integrated;][]{SuEtAl10, Dobler12LastLook, HuangEtAl13, HooperSlatyer13, FermiBubbles14} and edge \citep{Keshetgurwich17} energy spectra indicate that the FBs arose a few Myr ago as super-massive black hole (SMBH) energy injection, which must have been collimated \citep{MondalEtAl22} and nearly perpendicular to the Galactic plane \citep{SarkarEtAl23}. The resulting outflows drove strong, Mach $\gtrsim 5$ forward shocks \citep{Keshetgurwich17}, preferentially heating the ions, which remain largely non-equilibrated with the electrons \citep{Keshetgurwich18}.

These collisionless shocks were found to accelerate a flat, $p\equiv -d\ln n/d\ln E \simeq 2.1$ energy spectrum of cosmic-ray (CR) electrons (CREs) \citep{Dobler12LastLook, PlanckHaze13}, which subsequently undergo Kraichnan-like \citep{Kraichnan65, BerezinskiiEtAl90Book} diffusion of coefficient $D\propto E^{1/2}$ downstream \citep{Keshetgurwich17}.
Despite considerable research, the sub-X-ray diffuse spectrum was characterized only in a narrow, $20$--$60$ GHz band, and the \gama-ray spectrum was modelled non-self-consistently, invoking in particular unnaturally low, $\sim1$--$10$ TeV CR cutoffs \citep[\emph{cf.}][]{SherfEtAl17} to explain the observed $\epsilon \sim50\GeV$ photon break.
In addition, hadronic models show a $E<1\GeV$ spectrum harder than observed, whereas present leptonic models do not consistently model CRE cooling, invoking a power-law spectrum extending to very high energies corresponding to unrealistically young bubbles \citep[][]{CrockerAharonian11, Fujita13, YangEtAl13, FermiBubbles14}.

Figure \ref{fig:SpectFits} shows the specific $I_\epsilon$ \gama-ray and $I_\nu$ microwave intensities, where $\epsilon$ and $\nu$ are the photon energy and frequency.
The above arguments indicate that the \gama-rays arise from the inverse-Compton (IC) scattering of the ambient radiation field by the CREs.
Starlight photons of mean $\myEps\simeq 1\eV$ energy\footnote{We use $\myEps$ for seed photons and $\epsilon$ for up-scattered photons.} dominate the interstellar radiation budget above the Galactic center, out to a height of at least $5\kpc$ \citep[\eg][]{FermiBubbles14, PopescuEtAl17}.
The massive FB-driven outflows, not incorporated in typical radiation models \citep[which may underestimate starlight even near the Galactic center;  \eg][]{AjelloEtAl16}, have likely elevated the starlight and infrared energy densities considerably, out to the $\sim10\kpc$ height of the bubbles, by removing much of the ambient absorbers.
Hence, the $\epsilon \sim50\GeV$ break is naturally explained as the $\epsilon \simeq 3(m_e c^2)^2/16\myEps \simeq 50 \myEps_1^{-1} \GeV$, Klein-Nishina   \citep[KN; \eg][]{RybickiBook79} suppression of IC-scattered starlight, whereas the $\epsilon \gtrsim \mbox{few} \GeV$ softening is shown below to be the signature of a cooling break.
Here, $m_e$ is the electron mass, $c$ is the speed of light, and $\myEps_1\equiv \myEps/1 \eV$.
In such a model, the $\sim50\GeV$ break should become less pronounced at higher latitudes, where it is increasingly washed out by the spectrally-flat, IC-scattered cosmic microwave background (CMB) photons; indeed, such a trend is hinted by spatially resolved \citep{HooperSlatyer13} and edge \citep{Keshetgurwich17} FB spectra.

\section{\bf L\lowercase{eptonic model}}

\begin{figure}[h!]
    \centerline{\hspace{-0.cm}\epsfxsize=1\linewidth\epsfbox{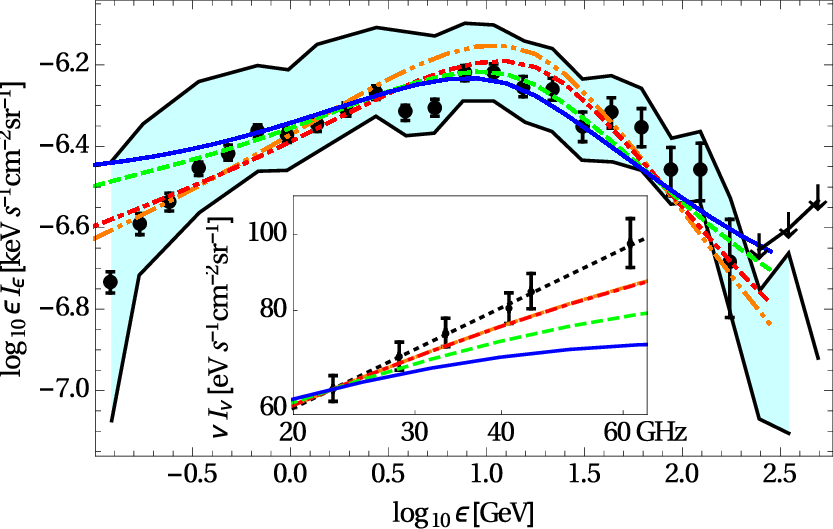}}
	\caption{\label{fig:SpectFits}
    Bulk \gama-ray \citep{FermiBubbles14} and (inset) microwave \citep{PlanckHaze13} FB spectra (black statistical error bars and upper-limit arrows, with cyan-shaded systematic uncertainty), synchrotron power-law fit ($I_\nu \propto \nu^{-0.57\pm0.06}$; dotted black curve), and one-zone leptonic models with $p=2.1$, $u_i=0.1\eV\cm^{-3}$, and additional parameters (with an increasing starlight intensity) $\{u_{s}\cm^3/\mbox{eV},t/\mbox{Myr},B/\mu\mbox{G}\}=\{0.5,10,1.45\}$ (solid blue),
    $\{1,5,2\}$ (dashed green),
    $\{2,2.4,2.7\}$ (dot-dashed red),
    and $\{3,1.8,3.1\}$ (double-dot-dashed orange).
}
\end{figure}

We assume that CREs are injected at the shock with index $p\simeq(r+2)/(r-1)$, thought to depend only on the shock compression ratio $r$ \citep{AxfordEtAl77, Krymskii77, Bell78, BlandfordOstriker78} if scattering is sufficiently isotropic \citep{keshet2020diffusive}, implying a shock velocity
\begin{equation}
\!\!\!\!\myv \simeq \sqrt{ \frac{2(p+2)\Gamma k_B T_u}{(2p+1-3\Gamma)\mathfrak{m}}} \simeq 1500\sqrt{\frac{T_{0.2}}{\myfp}}\km\se^{-1}\fin
\end{equation}
Here, $\mathfrak{m}\simeq 0.59m_p$ and $m_p$ are the mean and proton masses, $T_u\equiv 0.2 T_{0.2}\keV/k_B$ \citep{MillerBregman13, KataokaEtAl13, Keshetgurwich18} is the upstream temperature, $k_B$ is Boltzmann's constant, and we adopted an adiabatic index $\Gamma=5/3$ in the last, $p\simeq 2.1$ approximation where $q\equiv (p-2)/0.1$ approaches unity.
Such a $p\simeq 2.1$ spectrum reproduces the hard synchrotron fit in Fig.~\ref{fig:SpectFits} for non-cooled CREs, and the flat $\sim 3\mbox{--}30\GeV$ spectrum for cooled CREs.
The FB age for a simple, $R\propto t^{2/3}$ Primakoff scaling \citep{CourantFriedrichs48, Keller56, BernsteinBook80} of the $R\equiv 10 R_{10}\kpc$ FB height is then \citep{Keshetgurwich18}
\begin{equation} \label{eq:tPrim}
t \simeq \frac{2R}{3\myv} \simeq 4.4R_{10}\sqrt{\frac{q}{T_{0.2}}}\Myr\fin
\end{equation}

FB CREs cool by synchrotron and IC emission, thus softening to an index $p+1$ above the cooling-break energy
\begin{equation} \label{eq:Ec}
E_b \simeq \frac{3m^{2}_{e}c^3}{4\sigma_T t u} \approx 35\, u_3^{-1} t_3^{-1} \GeV \coma
\end{equation}
implying a synchrotron break at frequency
\begin{equation}\label{eq:SynBreak}
  \nu_b \simeq \frac{3 B e m_e c}{8 \pi (\sigma_T t u)^2} \simeq 51\, u_3^{-2} t_3^{-2} B_3 \GHz \coma
\end{equation}
and an IC-scattered starlight break at energy
\begin{equation}\label{eq:ICBreak}
  \epsilon_b \simeq \frac{3\myEps}{4} \left(\frac{m_e c}{\sigma_T t u}\right)^2
  \simeq 6.1\, u_3^{-2} t_3^{-2} \myEps_1 \GeV \coma
\end{equation}
where the energy density
\begin{equation}\label{eq:u}
  u \simeq u_s+u_i+u_c+\frac{B^2}{8\pi} \equiv 3\, u_3 \eV \cm^{-3}
\end{equation}
accounts respectively for the ambient $\myEps_s\simeq 1\eV$  starlight, $\myEps_i\simeq 10^{-2}\eV$ infrared \citep[\eg][]{FermiBubbles14}, $\myEps_c\simeq 10^{-3}\eV$ CMB, and magnetic, components.
Here, $t_3\equiv t/3\Myr$, $B_3\equiv B/3\muG$ is the normalized magnetic field, $e$ is the electron charge, and $\sigma_T$ is the Thomson cross-section.
The ratio between synchrotron emissivity at frequency $\nu$ from a non-cooled, $n\propto E^{-p}$ CRE distribution, and starlight-IC emissivity at energy $\epsilon$ from the cooled CREs, is
\begin{eqnarray} \label{FluxRatio}
\Phi\equiv \frac{\nu j_\nu}{\epsilon j_\epsilon} & \simeq & \phi(p) m_e^{-\frac{p+3}{2}} e^{\frac{p+5}{2}} c^{-\frac{p+7}{2}} \frac{ut}{u_s} B^{\frac{p+1}{2}} \nu^{\frac{3-p}{2}} \left(\frac{\epsilon}{\myEps}\right)^{\frac{p-2}{2}} \nonumber \\
    & \simeq & 0.043 \frac{u}{u_s} t_3 B_3^{\frac{p+1}{2}} \nu_{10}^{\frac{3-p}{2}}
    \left(\frac{\epsilon_{10}}{0.46\myEps_1}\right)^{\frac{p-2}{2}} \coma
\end{eqnarray}
where we defined $\nu_{10}\equiv \nu/10\GHz$ and $\epsilon_{10}\equiv\epsilon/10\GeV$;
the dimensionless function $\phi(p)$ is given in Appendix \S\ref{sec:Leptonic}.
Adopting the $u_i+u_c\ll u_s$ limit (henceforth), the ratio between radio and $\gtrsim 10\GeV$ \gama-ray intensities should approximately be given by Eq.~\eqref{FluxRatio}.

FB observations indicate a fairly uniform CR distribution within the bubbles \citep{SuEtAl10, FermiBubbles14}, so $t$, $B$, $u$, and $u_s$ can be determined in a one-zone approximation from Eqs.~\eqref{eq:SynBreak}--\eqref{FluxRatio}, given $p$, $\myEps$, $\nu_b$, $\epsilon_b$, and $\Phi$ measurements.
An $\alpha\equiv d\ln j_\nu/d\ln\nu \simeq -0.55$ bulk microwave spectral index extending out to $\gtrsim 60\GHz$ \citep[][and Fig.~\ref{fig:SpectFits}]{PlanckHaze13} implies that $p=1-2\alpha\simeq 2.1$.
Hence, the bulk and edge $\gamma$-ray spectra \citep{SuEtAl10, FermiBubbles14, Keshetgurwich17} are consistent with an $\epsilon_b\equiv 3 \epsilon_{b,3} \GeV \simeq 3\GeV$ cooling break, we may use the observed ratio $\Phi\equiv 0.1\Phi_{0.1}\simeq 0.12$ between (henceforth) the $\epsilon=10$ Gev \citep{FermiBubbles14} and $\nu=23\GHz$ \citep{PlanckHaze13} fluxes, and, taken at face value, $\nu_b>60\GHz$.
Equations \eqref{eq:ICBreak}--\eqref{FluxRatio} then yield the plausible
$u_3t_3\simeq 1.4(\myEps_1/\epsilon_{b,3})^{1/2}$,
$t_3B_3^{1.55}\simeq1.5\myEps_1^{0.05}\Phi_{0.1}u_s/u$,
and
$u_s/u\simeq 1-0.074B_3^2u_3^{-1}$.
However, the putative lower $\nu_b$ limit would then lead, via Eq.~\eqref{eq:SynBreak}, to the unrealistic inequalities
$B_3\gtrsim2.4\epsilon_{b,3}^{-1}\myEps_1$,
$t_3\lesssim0.40\myEps_1^{-3/2}\epsilon_{b,3}^{1.55}\Phi_{0.1}u_s/u$,
and $u_s\gtrsim 10\myEps_1^2\epsilon_{b,3}^{-2.05}\Phi_{0.1}^{-1}\eV\cm^{-3}$.

\section{\bf M\lowercase{issing microwave cooling break}}

As the FBs cannot be basked in so much starlight, and are unlikely to be so young, the model is plausible only in the presence of a $\nu_b<60\GHz$ cooling break, unidentified until now due to an additional microwave component, most likely dust.
Indeed, replacing the spectral break Eq.~\eqref{eq:SynBreak}
by the hydrodynamic Eq.~\eqref{eq:tPrim} fixes the model parameters
\footnote{We neglect a weak, $\myEps_1^{q/(30+q)}$ dependence of $B$ and $\nu_b$.}
$t\simeq4.4R_{10}(\myfp/T_{0.2})^{1/2}\Myr$,
$u\simeq 2.9R_{10}^{-1}(T_{0.2}\myEps_1/ \epsilon_{b,3}\myfp)^{1/2}\eV\cm^{-3}$,
and
$B\simeq 3.1(T_{0.2}/\myfp)^{Q/3}(\Phi_{0.1}u_s/R_{10}u)^{2Q/3}\muG$,
where we defined $Q\equiv (1+q/30)^{-1}\simeq 1$.
Equation \eqref{eq:SynBreak} then predicts a spectral break at frequency
\begin{equation}\label{Eq:HydroNub}
 \nu_b\simeq 26 \frac{\epsilon_{b,3}}{\myEps_1}
 \left(\frac{T_{0.2}}{\myfp}\right)^{Q/3}\left(\frac{\Phi_{0.1}u_s}{R_{10}u}\right)^{2Q/3}\GHz \fin
\end{equation}

Figure \ref{fig:SpectFits} demonstrates a few one-zone leptonic-model fits to the bulk \gama-ray spectrum, constrained also by the $\nu\simeq 23\GHz$ \emph{WMAP} data point.
Here, we fix $p=2.1$ and $u_i=0.1\eV\cm^{-3}$, and vary the other parameters; additional model fits are provided in Appendix \S\ref{sec:Leptonic}.
Interestingly, for the best-fit parameters, we find that the cooling break and KN turnaround approximately merge to yield a stronger spectral break.
While the $u_{s}=0.5\eV\cm^{-3}$ fit marginally agrees, within statistical and systematic uncertainties, with the \gama-ray data, much better fits are obtained for $1\eV\cm^{-3}\lesssim u_{s}\lesssim 3\eV\cm^{-3}$.
Such starlight energy densities are higher than usually assumed at $R\gtrsim 4\kpc$ heights, but are plausible as the mean starlight inside the FB outflows, given conservative $u_s(R=2\kpc)\simeq 1.4\eV\cm^{-3}$ estimates \citep{FermiBubbles14}, as discussed above and in \S\ref{sec:Discussion}.
For such $u_s$ values, the \gama-ray fit is surprisingly good, considering the oversimplified, one-zone approximation, and can be arbitrarily improved if this assumption is relaxed.
In contrast, the microwave fit is poor in all our successful \gama-ray models, as they require a $\nu_b<40\GHz$ cooling break.
These quantitative conclusions are in line with the above qualitative estimates, and are not sensitive to our assumptions; in particular, we obtain similar results for different $2.0\leq p\lesssim 2.2$ CRE spectra, different $u_i<0.2u_s$ prescriptions, and two-zone models.
For $p\gtrsim 2.3$, the \gama-ray spectral fit becomes unacceptable for plausible $t$ and $u_s$ values, as shown in Appendix \S\ref{sec:Leptonic}.

\section{\bf B\lowercase{ubble-edge analysis}}

Next, we analyze the FB edges, identified by gradient filters \citep{Keshetgurwich17}, and bin the data parallel to the edge in the same method used to measure the \gama-ray \citep{Keshetgurwich17} and X-ray \citep{Keshetgurwich18} spectra in the near downstream.
Here, we study lower frequencies, ranging from $45\MHz$ radio to $60\,\mu\mbox{m}$ infrared, focusing in particular on cleaned microwave \emph{WMAP} and \emph{Planck} maps with point sources masked and the CMB subtracted, as described in Appendix \S\ref{sec:Planck}.
Figure \ref{fig:RGB} demonstrates (RGB colors) the \gama-ray, infrared, and microwave sky in and around the FBs, and outlines (white contours) the north and south sectors used in our analysis.
These sectors, crossing the FB edge at the top of each bubble, were chosen to contain large upstream and downstream regions with minimal extended contamination.

\begin{figure}[h!]
    \centerline{
    \includegraphics[width=1\linewidth,trim={0cm 0cm 0cm 0cm},clip]{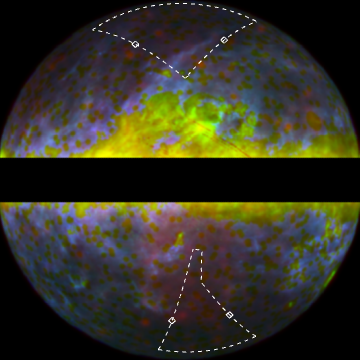}
    }
	\caption{\label{fig:RGB}
    \!\!False-color image combining \emph{Fermi}-LAT $3$--$30\GeV$ (red; see Appendix \S\ref{sec:Fermi}), \emph{AKARI}-FIS WideL $140\,\mu\mbox{m}$ \citep[green;][]{AKARI15}, and \emph{Planck} $143\GHz$ (blue) in an orthographic projection; the bright, $|b|\lesssim5^\circ$ Galactic plane was masked.
    Our north and south sectors are highlighted (dashed white boundaries with diamonds where they intersect the FB edges).
	}
\end{figure}

\begin{figure*}[t!]
    \centerline{\epsfxsize=0.47\linewidth\epsfbox{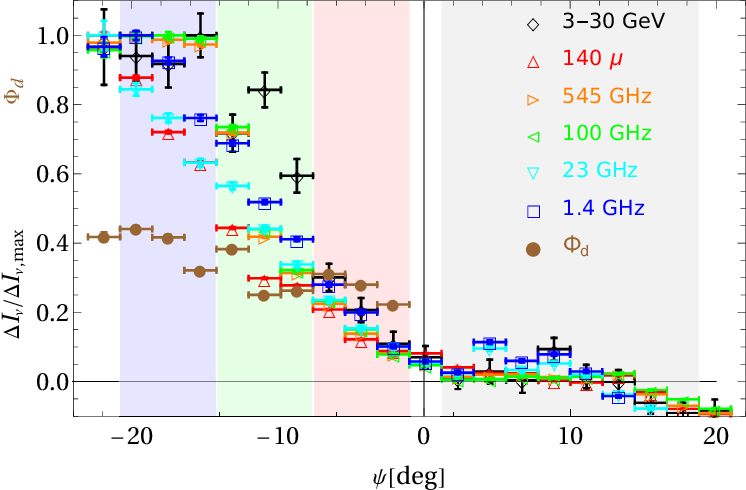}\hspace{0.3cm}\epsfxsize=0.49\linewidth\epsfbox{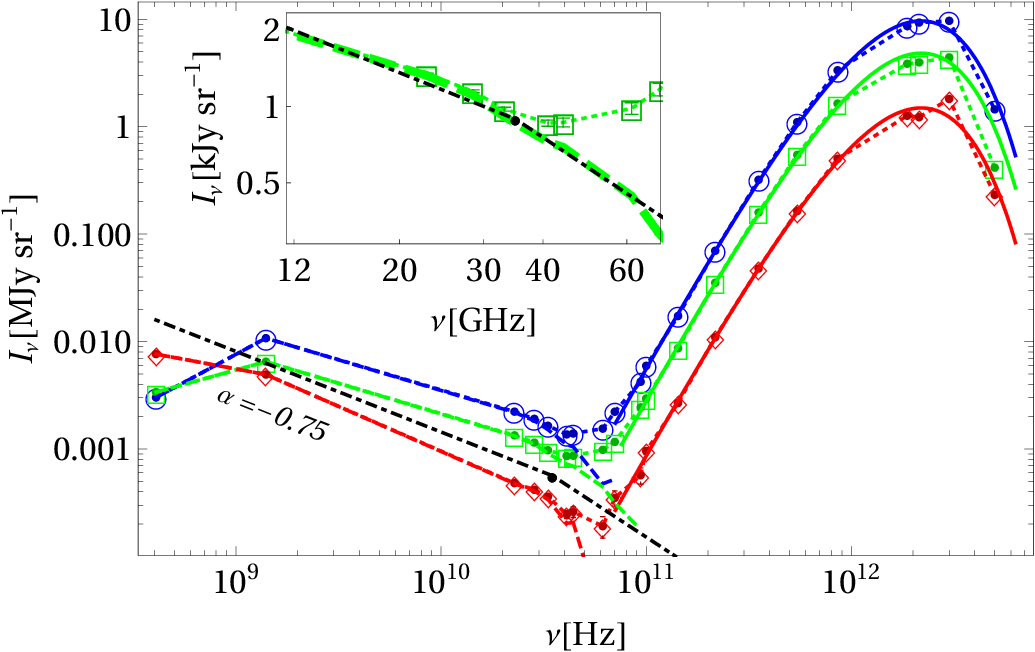}}
    \vspace{0.2cm}
    \centerline{\epsfxsize=0.47\linewidth\epsfbox{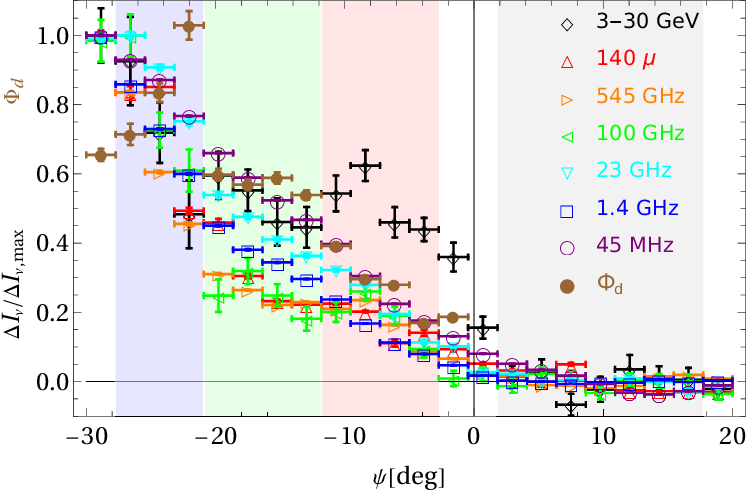}\hspace{0.3cm}\epsfxsize=0.49\linewidth\epsfbox{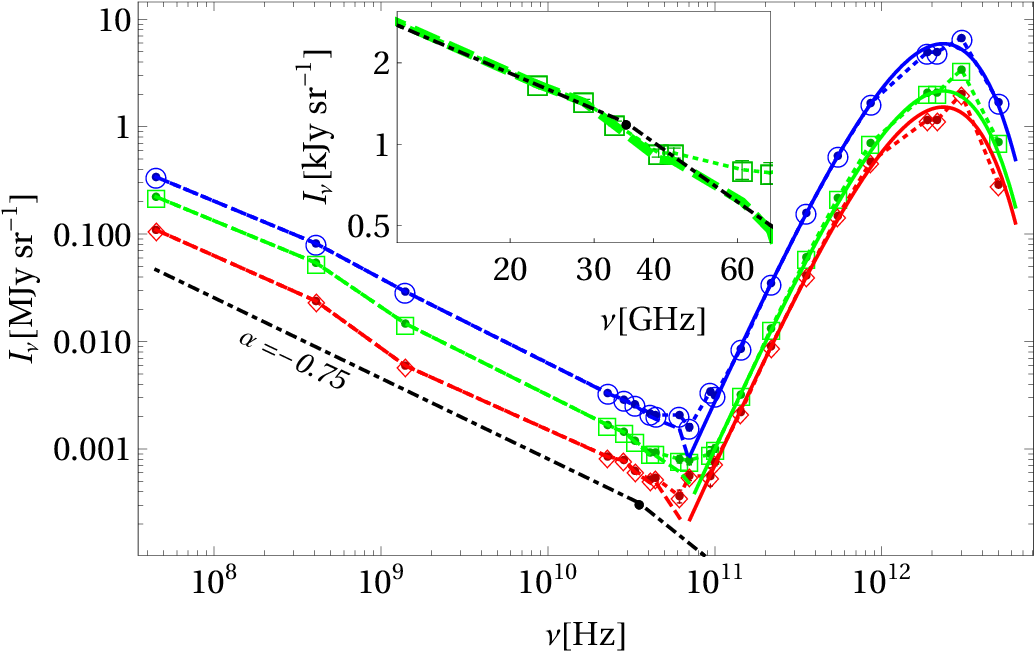}}
	\caption{\label{fig:Edges}
   Spectral analysis of north (top panels) and south (bottom) FB edges.
   Left panels show (error bars) the profiles of select (see legend) channels (normalized by their maxima) and the $23\GHz$ to $10\GeV$ brightness ratio $\Phi_d$, as a function of the angular distance $\psi$ from the edge, after subtracting the foreground based on the upstream ($\psi>0$) gray-shaded region.
   Right panels show (symbols enclosing error bars, with dotted lines to guide the eye) the spectrum of excess emission in the near (red diamonds), mid (green squares), and far (blue circles) downstream regions designated by the respective color-shaded regions in the left panels.
   Also shown are the synchrotron profiles (dashed curves) obtained by subtracting the respective best-fitted dust models (see text; solid curves) from the downstream excess, and a synchrotron model (of arbitrary normalization; labeled dot-dashed black) of $p=2$ CRE injection, Kraichnan diffusion, and a $\nu_b=35\GHz$ cooling break (black disk); the insets focus on the cooling-break region in the mid-downstream.
   Data are extracted from $45\MHz$ \citep{AlvarezEtAl97, MaedaEtAl99}, reprocessed $408\MHz$ \citep{Haslam,Remazeilles2015}, $1.4\GHz$ \citep{Reich82, ReichReich86, ReichEtAl01}, \emph{WMAP} and \emph{Planck} microwave, $160\,\mu\mbox{m}$ and $140\,\mu\mbox{m}$ AKARI \citep{AKARI15}, $100\,\mu\mbox{m}$ and $60\,\mu\mbox{m}$ IRIS \citep{IRIS05}, and $3$--$30\GeV$ \emph{Fermi}-LAT maps.
   	}
\end{figure*}

Figure \ref{fig:Edges} presents the spectrum of the downstream regions of both north (upper panels) and south (bottom panels) FBs, obtained by binning the data parallel to the edge and subtracting the upstream-based foreground and background (henceforth foreground).
The spatial distribution is demonstrated (left panels) for select channels, all found to sharply brighten as one crosses downstream, facilitating a measurement of the spectrum (right panels) at different outward-oriented angular distances $\psi$ from the edge.
While the signal strengthens downstream with increasing $|\psi|$, the spectral shape is found to be robust as a function of $\psi$, and quite similar in the north and south bubbles, substantiating the validity of the measurement.
As expected from previous studies, the north bubble shows more surrounding substructure, both upstream (Loop-I) and downstream (mainly at low frequencies); it also presents a more gradual \gama-ray jump, possibly attributed in part to edge misalignment.

The cleaner, south bubble provides a good measurement of the synchrotron signal, found for the first time to persist down to frequencies as low as $45\MHz$.
In the present method, the spectral index can be accurately determined to be $\alpha=-0.77\pm0.03$ downstream (see Table \ref{tab:Reg}).
This index is softer by an $\eta=0.20\pm0.07$ offset with respect to the $\alpha=-0.57\pm0.06$ bulk synchrotron index (Fig.~\ref{fig:SpectFits}), consistent with both Kraichnan and Kolmogorov diffusion of escaping CREs.
A similar but slightly stronger softening was identified in the IC \gama-ray emission from the edge, and was interpreted as arising from CRE diffusion \citep{Keshetgurwich17}; its detection in radio validates this diffusive model.

A pronounced hard signal is found to dominate the downstream spectrum above $\sim100\GHz$, extending up to an exponential turnaround in the infrared. The signal is well-fitted by an $I_\nu\propto \nu^{7/2}/(e^{h\nu/k_B T}-1)$ spectrum (solid curves in Fig.~\ref{fig:Edges}), and is thus identified as thermal dust emission, of temperature $T\simeq 29\pm1\mbox{ K}$ ($33\pm1\mbox{ K}$) in the north (south) FB (see Table \ref{tab:Reg}).
Such temperatures, already hinted by coincident hot dust regions in Planck maps, may underestimate the true postshock $T$ due to the inaccurate subtraction of $\lesssim 20\mbox{ K}$ \citep{Planck16_Disentangling} foreground dust.
As the left panels of Fig.~\ref{fig:Edges} show, the radial profiles of dust emission are nearly identical to those of IC and especially synchrotron emission, in both bubbles.
We conclude that this dust component is directly associated with the FB shock, probably due to dust entrainment or formation in the postshock region, in resemblance of AGB stars, novae, and core-collapse supernovae.
Indeed, Fig.~\ref{fig:RGB} shows some of the dust emission (green and blue) extending to the top of the \gama-ray (red) bubbles.
A dust FB component should therefore be included in template analyses that aim at measuring the microwave spectrum.

\begin{deluxetable}{llccc}
  \tablewidth{0pt}
  \tablecaption{Spectral synchrotron+dust model fit. \label{tab:Reg}}
  \tablehead{
  \colhead{Bubble} & \colhead{Region$^a$} &  \colhead{$\alpha$} &  \colhead{$\nu_b$} & \colhead{$T$} \\
  \colhead{} & \colhead{} & \colhead{} & \colhead{GHz} & \colhead{K}
  }
  \startdata
\colrule
North & Near & $-0.88\pm0.10$ & $27.8\pm14.4$ & $30.0\pm0.8$ \\
 & Mid & $-0.59\pm0.08$ & $24.8\pm9.7$ & $27.8\pm0.5$ \\
 & Far & $-0.59\pm0.06$ & $23.6\pm7.2$ & $29.6\pm0.4$ \\
\colrule
South & Near & $-0.78\pm0.03$ & $34.7\pm12.5$ & $31.5\pm0.9$ \\
 & Mid & $-0.80\pm0.03$ & $29.3\pm8.6$ & $33.9\pm0.8$ \\
 & Far & $-0.75\pm0.02$ & $37.4\pm7.6$ & $32.5\pm0.5$ \\
  \enddata
  \tablenotetext{a}{The downstream regions are defined in Fig.~\ref{fig:Edges}; in north bubble regions, $\nu<$ GHz data were excluded from the fit due to interfering substructure.}
\end{deluxetable}\label{t:Reg}

\section{\bf J\lowercase{oint bulk and edge model}}

The unexpected, hard microwave dust-emission downstream explains in part why a cooling break was not identified until now.
Indeed, Fig.~\ref{fig:Edges} shows that if the dust model is subtracted from the data, the remaining synchrotron component (dashed curves) presents a spectral break (insets) around $\sim35\GHz$, as anticipated above.
In accordance, fitting a superposition of synchrotron emission of index $\alpha$ and a frequency $\nu_b$ cooling break, and thermal dust emission of temperature $T$, as summarized in Table \ref{tab:Reg}, gives $\nu_b=34\pm5\GHz$ in the southern bubble.
Curiously, both \emph{WMAP} \citep{Dobler12LastLook} and joint \emph{WMAP} and \emph{Planck} \citep{PlanckHaze13} template-based analyses found a similar bulk spectral break in the total synchrotron signal, but attributed it to combined hard+soft foregrounds rather than to the bubbles.

With this spectral break, solving Eqs.~\eqref{eq:SynBreak}--\eqref{FluxRatio} in the downstream region, where $p=1-2\alpha\simeq 2.5$, yields
\begin{equation}
 t\simeq 5.2\myEps_1^{-3/2}\left(\frac{\epsilon_{b,3}}{\nu_{b,35}}\right)^{7/4}
 \frac{u_s}{u}\Phi_{0.2}\Myr \coma
 \end{equation}
\begin{equation}
 u_s\simeq 2.5\myEps_1^{2}\epsilon_{b,3}^{-9/4}\nu_{b,35}^{7/4}\Phi_{0.2}^{-1}\eV\cm^{-3} \coma
\end{equation}
\begin{equation}
B\simeq 4.2\myEps_1\epsilon_{b,3}^{-1}\nu_{b,35}\muG \coma
\end{equation}
and
\begin{equation}
u/u_s=1+0.17(\epsilon_{b,3}\nu_{b,35})^{1/4}\Phi_{0.2} \coma
\end{equation}
where we defined $\nu_{b,35}\equiv \nu_{b}/35\GHz$
and a downstream $\Phi_{0.2}\equiv \Phi_d/0.2$ (see Fig.~\ref{fig:Edges}).
While $\Phi_d\simeq 0.2$ remains fairly uniform in the north, it increases inward beyond $\psi\sim-10^\circ$ in the south; however, our method is increasingly sensitive to foreground variations at large $|\psi|$.

While the local spectral measurement near the edge does not require strong assumptions concerning extended foregrounds, and does not suffer from template-removal systematics \citep{Keshetgurwich17, Keshetgurwich18} which can be highly detrimental \citep[\eg][]{KeshetEtAl04_EGRB}, it does show larger statistical errors than the volume-integrated, bulk spectrum.
Our results indicate that this bulk spectrum too must show a cooling break, which would fix the FB parameters even without using the hydrodynamic Eq.~\eqref{eq:tPrim}.
Requiring similar FB ages inferred from edge and bulk spectra suggests a $\nu_{\mbox{\tiny{bulk}}}\simeq 23\GHz$ cooling break in the latter, unidentified so far due to modelling systematics and unaccounted FB dust.
We then obtain for the bulk
$t\simeq 5.2\myEps_1^{-3/2}(\epsilon_{b,3}/\nu_{b,23})^{1.55}(u_s/u)\Phi_{0.1}\Myr$,
$u_s\simeq 2.5\myEps_1^{2}\epsilon_{b,3}^{-2.05}\nu_{b,23}^{1.55}\Phi_{0.1}^{-1}\eV\cm^{-3}$,
$u/u_s \simeq 1+0.077\nu_{b,23}^{0.45}\Phi_{0.1}$, and
\begin{equation}
B\simeq 2.8\myEps_1\epsilon_{b,3}^{-1}\nu_{b,23}\muG \coma
\end{equation}
where we defined $\nu_{b,23}\equiv \nu_{\mbox{\tiny{bulk}}}/23\GHz$.
Note that while a $\nu_{\mbox{\tiny{bulk}}} < 60\GHz$ cooling break would modify the previously reported $\alpha\simeq 0.55$ synchrotron spectrum, a $p\simeq 2.1$ CRE spectrum is nevertheless needed for a self-consistent model, as well as for an acceptable \gama-ray fit (see Appendix \ref{sec:Leptonic}).

\section{\bf D\lowercase{iscussion}}
\label{sec:Discussion}

We conclude that the spectra of both north and south FB edges, and of the bulk, both with and without hydrodynamic considerations, produce consistent results, thus providing strong constraints on the FBs and their environment.
The results indicate younger FBs than obtained in early spectral models, an enhanced magnetic field near the edges, and strong and fairly uniform radiation fields throughout the bubbles.
Indeed, while we cannot directly measure the ambient starlight intensity, Fig.~\ref{fig:Edges} shows an infrared $\nu u_\nu\simeq 0.1\eV\cm^{-3}$ peak  at high-latitudes, very similar to its estimated value near the Galactic plane \citep[\eg~$\nu u_\nu\simeq 0.15\eV\cm^{-3}$ at $R=2\kpc$;][]{FermiBubbles14}.

The success of our oversimplified, one-zone bulk model, which attributes the synchrotron and IC signals to the same pure power-law CRE injection, invalidates the justification for previous hadronic or leptonic models, which invoked ad-hoc cutoffs on the CR spectrum.
Without ad-hoc assumptions, future better measurements of the broadband, radio to \gama-ray signal would robustly constrain the CRE distribution and diffusion, coincident radiation fields, magnetic fields, and FB dynamics and energetics.
The edge spectra presented in this work validate the bulk model, directly confirming the CRE softening expected from diffusion \citep{Keshetgurwich17}, the anticipated microwave cooling break, and, indirectly, the persistence of strong radiation fields throughout the FBs.

\vspace{0.3cm}
We thank K.C. Hou, O. Freifeld,  E. Waxman, and R. Crocker for useful discussions.
This research was supported by the Israel Science Foundation (Grant No. 2126/22), by the IAEC-UPBC joint research foundation (Grant No. 300/18), and by the Ministry of Science, Technology \& Space, Israel.


\appendix
\twocolumngrid

\section{\bf L\lowercase{eptonic model}}
\label{sec:Leptonic}

\begin{figure}[t!]
    \centerline{\hspace{-0.cm}\epsfxsize=1\linewidth\epsfbox{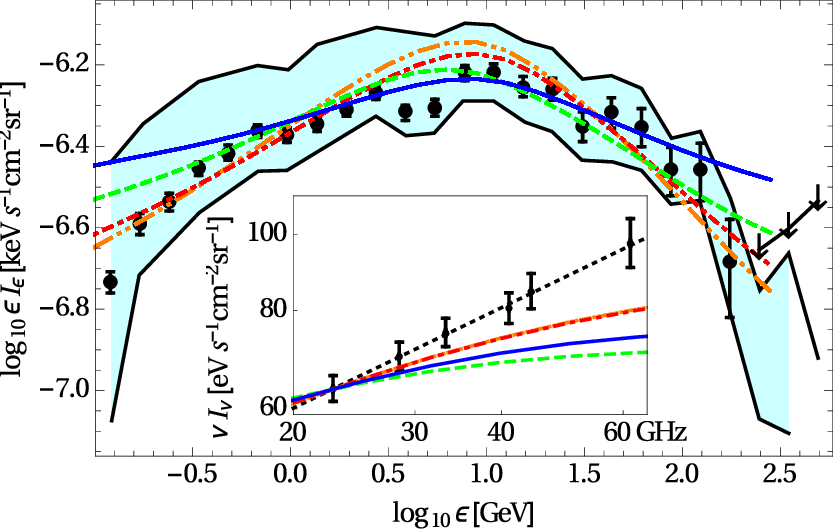}}
	\caption{\label{fig:SpectFitP20}
    Same as Fig.~\ref{fig:SpectFits}, but for $p=2$ and model parameters $\{u_{s}\cm^3/\mbox{eV},t/\mbox{Myr},B/\mu\mbox{G}\}=
    \{0.5,10,1.7\}$ (solid blue),
    $\{1,8,2\}$ (dashed green),
    $\{2,3.6,2.6\}$ (dot-dashed red),
    and $\{3,2.7,3\}$ (double-dot-dashed orange).
}
\end{figure}

\begin{figure}[h!]
    \centerline{\hspace{-0.cm}\epsfxsize=1\linewidth\epsfbox{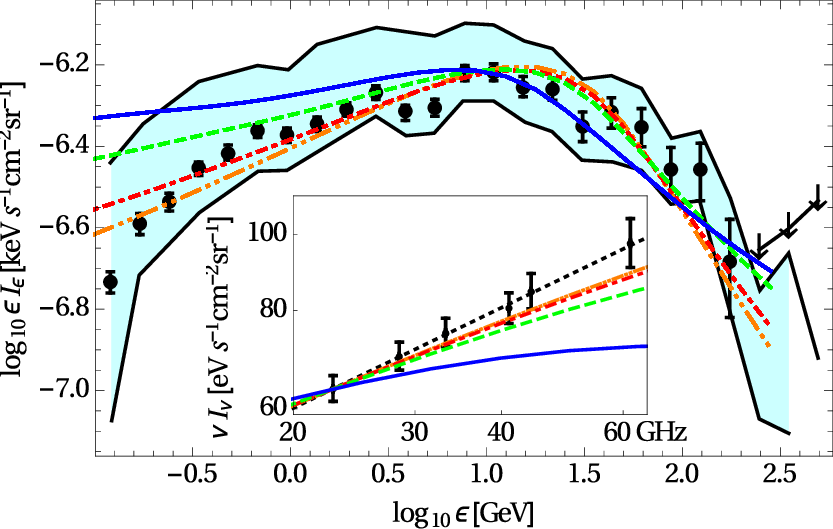}}
	\caption{\label{fig:SpectFitP22}
    Same as Fig.~\ref{fig:SpectFits}, but for $p=2.2$ and model parameters $\{u_{s}\cm^3/\mbox{eV},t/\mbox{Myr},B/\mu\mbox{G}\}=
    \{0.5,8,1.45\}$ (solid blue),
    $\{1,3.2,2\}$ (dashed green),
    $\{2,1.6,2.8\}$ (dot-dashed red),
    and $\{3,1.1,3.4\}$ (double-dot-dashed orange).
}
\end{figure}

\begin{figure}[h!]
    \centerline{\hspace{-0.cm}\epsfxsize=1\linewidth\epsfbox{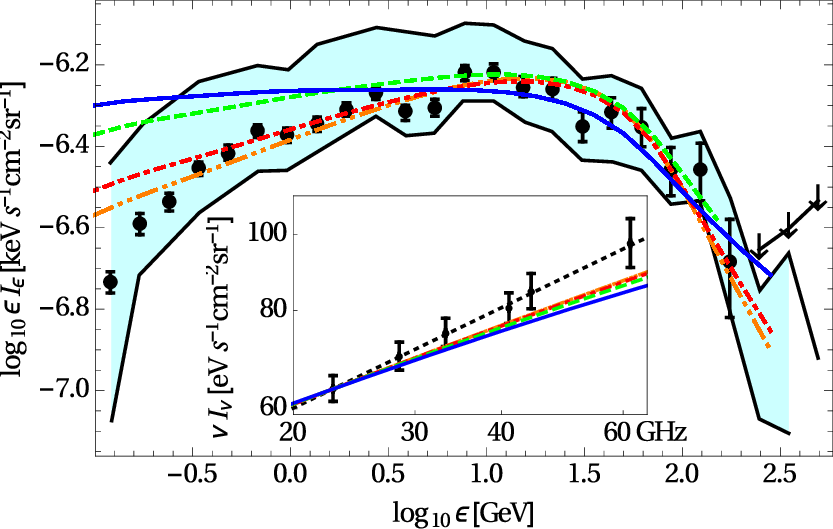}}
	\caption{\label{fig:SpectFitP23}
    Same as Fig.~\ref{fig:SpectFits}, but for $p=2.3$ and model parameters $\{u_{s}\cm^3/\mbox{eV},t/\mbox{Myr},B/\mu\mbox{G}\}=
    \{0.5,3.3,1.7\}$ (solid blue),
    $\{1,1.8,2.1\}$ (dashed green),
    $\{2,1,2.9\}$ (dot-dashed red),
    and $\{3,0.7,3.5\}$ (double-dot-dashed orange).
}
\end{figure}

\begin{figure}[b!]
    \centerline{\hspace{-0.cm}\epsfxsize=1\linewidth\epsfbox{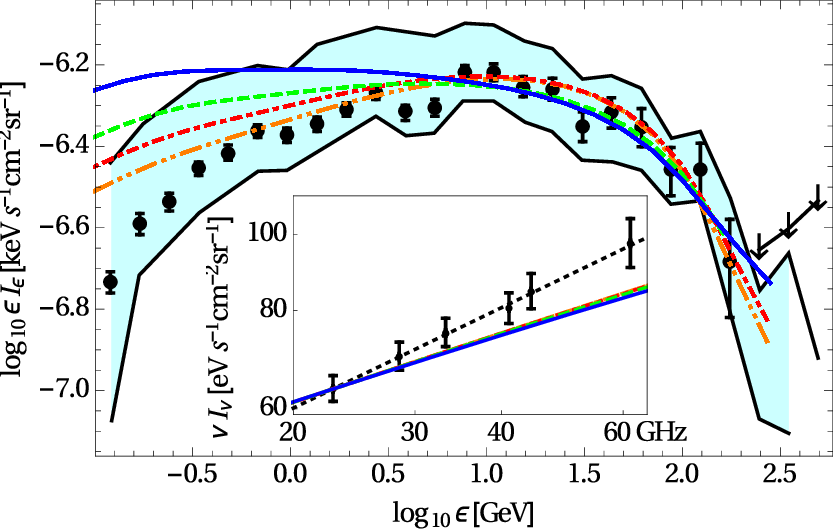}}
	\caption{\label{fig:SpectFitP24}
    Same as Fig.~\ref{fig:SpectFits}, but for $p=2.4$ and model parameters $\{u_{s}\cm^3/\mbox{eV},t/\mbox{Myr},B/\mu\mbox{G}\}=
    \{0.5,2.2,1.7\}$ (solid blue),
    $\{1,1.2,2.2\}$ (dashed green),
    $\{2,0.7,2.8\}$ (dot-dashed red),
    and $\{3,0.5,3.4\}$ (double-dot-dashed orange).
}
\end{figure}

In our one-zone leptonic model, FB \gama-ray emission is dominated the the IC scattering of starlight.
For simplicity, we approximate the integrated CR population as injected at a constant rate, so its evolved spectrum is a simple broken power-law.
Equations \eqref{eq:Ec}--\eqref{FluxRatio} follow directly \citep[\eg][]{RybickiBook79}, where the dimensionless prefactor
\begin{eqnarray}\label{eq:Cp}
\phi & \equiv & \frac
{2^{\frac{3}{2}}\pi^{\frac{3 p + 2}{2}}(6 + p)(4 + p)\Gamma\left(\frac{p}{4} - \frac{1}{12}\right)
    \Gamma\left(\frac{19}{12} + \frac{p}{4}\right)\Gamma\left(\frac{p + 1}{4}\right)}
    {3^2 80^{\frac{p}{2}} (16 + 6 p + p^2) \Gamma\left(\frac{p+2}{2}\right)
    \Gamma\left(\frac{7 + p}{4}\right)\zeta\left(\frac{6 + p}{2}\right)\zeta(3)^{\frac{p-2}{2}}} \nonumber \\
    & \simeq & 30.5\times 2.7^{-1.6 p} \,\mbox{(approximation for $2<p<3$)}
\end{eqnarray}
of the ratio $\Phi$ between non-cooled synchrotron and cooled IC spectra is obtained after averaging over photons in a seed distribution of temperature $T$,
\begin{equation}\label{eq:EAvg}
  \epsilon_0 \equiv \langle \epsilon\rangle_T =
  \frac{\pi^4 k_B T}{30\zeta(3)} \simeq 2.7k_B T\coma
\end{equation}
and averaging the pitch angle $\alpha$ over an isotropic distribution,
\begin{equation}\label{eq:PitchAvg}
  \langle \sin(\alpha)^{\frac{p+1}{2}}\rangle = \frac{\pi^{\frac{1}{2}}\Gamma\left(\frac{p+5}{4}\right)}{2\Gamma\left(\frac{p+7}{4}\right)}
  \simeq \frac{14}{17}-\frac{p}{19}\coma
\end{equation}
approximated for $2<p<3$.
Here, $\zeta(s)$ is the Riemann Zeta function and $\Gamma(z)$ is the Euler gamma function.
Our numerical spectral computations, incorporating the KN suppression, follow Refs. \citep{Jones68,AharonianAtoyan81};
results are shown for $p=2.0$ and $p=2.2$ in Figs.~\ref{fig:SpectFitP20} and \ref{fig:SpectFitP22}.
For $p\gtrsim 2.3$, the \gama-ray fit becomes increasingly worse and requires unreasonably young FBs and intense starlight, as illustrated in Figs.~\ref{fig:SpectFitP23} and \ref{fig:SpectFitP24}.

\section{\bf M\lowercase{icrowave sky maps}}
\label{sec:Planck}

We combine \emph{Planck} mission and  \emph{Wilkinson Microwave Anisotropy Probe} (\emph{WMAP}) data to produce the microwave maps for the analysis. We use the full mission, bandpass-leakage corrected, $28.5,\ 44.1,\ 70.3,\ 100.0,$ and $143.0$ GHz maps from \emph{Planck}\footnote{See \url{https://pla.esac.esa.int/\#home}}, and the nine-year, deconvolved maps of the $22.8,\ 33.2,\ 41.0,\ 61.4,$ and $94.0$ GHz bands from \emph{WMAP}\footnote{See \href{https://lambda.gsfc.nasa.gov/product/map/dr5/maps_deconv_band_r9_i_9yr_get.cfm}{https://lambda.gsfc.nasa.gov/product/map/dr5}\\
\href{https://lambda.gsfc.nasa.gov/product/map/dr5/maps_deconv_band_r9_i_9yr_get.cfm}{/maps\_deconv\_band\_r9\_i\_9yr\_get.cfm}
}.

Binning along the FB edge removes most foregrounds, but to obtain better results, we first subtract the CMB contribution, estimated using the
Planck HFI internal linear combination \cite[PILC;][]{PlanckHaze13}.
This map is constructed by a superposition of 143--545 GHz maps, designed to spectrally isolate the CMB from the dust template of \citet{FDS99}.

We mask the point sources from \emph{WMAP}'s nine-year point-source catalog and \emph{Planck}'s second catalogue of point sources (PCCS2) in the relevant bands, and the following large-scale structures: the Small and Large Magellanic Clouds, M31, NGC5090, NGC5128, Orion-Barnard Loop, and $\zeta$ Oph.
We also mask pixels where dust extinction at H$\alpha$ frequencies is larger than 1 mag,
and pixels where the H$\alpha$ intensity is greater than 10 Rayleigh.
All maps are masked at the highest, $N_{\mbox{\tiny{side}}}=2048$ HEALPix \citep{GorskiEtAl05} resolution available.

\section{\bf F\lowercase{ermi-}LAT \lowercase{sky map}}
\label{sec:Fermi}

We use the archival Pass-8 LAT data from the Fermi Science Support Center (FSSC)\footnote{See \url{http://fermi.gsfc.nasa.gov/ssc}}, and the Fermi Science Tools (version \texttt{v10r0p5}).
Weekly all-sky files are used, spanning weeks $9$ through $789$ for a total of $781$ weeks ($\sim15$ years), with ULTRACLEANVETO class photon events.
A $90^\circ$ zenith angle cut is applied to avoid CR-generated $\gamma$-rays from the Earth's atmospheric limb, according to the appropriate
FSSC Data Preparation recommendations.
Good time intervals are selected using the recommended expression \texttt{(DATA\_QUAL==1) and (LAT\_CONFIG==1)}.
Point-source contamination is minimized by masking pixels within the $95\%$ total-event containment area of each point source in the LAT fourth source catalog \cite[4FGL;][]{AbdollahiEtAl20}.

\bibliography{FermiBubbles}

\end{document}